# Futuristic methods in virus genome evolution using the Third-Generation DNA sequencing and artificial neural networks

Hyunjin Shim

**Abstract**

The Third-Generation in DNA sequencing has emerged in the last few years using new technologies that allow the production of long-read sequences. Applications of the Third-Generation sequencing enable real-time and on-site data production, changing the research paradigms in environmental and medical sampling in virology. To take full advantage of large-scale data generated from long-read sequencing, an innovation in the downstream data analysis is necessary. Here, we discuss futuristic methods using machine learning approaches to analyze big genetic data. Machine learning combines pattern recognition and computational learning to perform predictive and exploratory data analysis. In particular, deep learning is a field of machine learning that is used to solve complex problems through artificial neural networks. Unlike other methods, features can be learned using neural networks entirely from data without manual specifications. We discuss the future of 21st-century virology by presenting futuristic approaches for virus studies using real-time data production and on-site data analysis with the Third-Generation Sequencing and machine learning methods. We first introduce the basic concepts in conventional statistical models and methods in virology, building gradually into the necessity of innovating the downstream data analysis to meet the advances in sequencing technologies. We argue that artificial neural networks can innovate the downstream data analysis, as they can learn from big datasets without model assumptions or feature specifications, as opposed to the current data analysis in bioinformatics. Furthermore, we discuss how futuristic methods using artificial neural networks combined with long-read sequences can revolutionize virus studies, using specific examples in supervised and unsupervised settings.

# 1 Table of Contents

# 1. Virus as Model Organism

Viruses are the most abundant biological entities on Earth, which can infect all types of life forms, and only replicate inside cellular organisms. Viruses exploit diverse replication-expression strategies, whereas cellular organisms only use double-stranded DNA as the replicating form and single-stranded RNA as the transcribed form [1]. Viruses also display all conceivable genome architectures – single or double-stranded, linear or circular, DNA or RNA, monopartite or multipartite – with sizes spanning three orders of magnitude [1]. Many viral pathogens exhibit complex evolutionary trajectories during transmission between hosts and diversification within them: they may experience strong inter-host population bottlenecks and rapid intra-host population growth, as well as intense selective pressures due to the host immunity and drug therapy [2], [3]. These phenomena have been investigated in a range of human pathogens impacting global health, including Human Immunodeficiency virus (HIV) [4] and Influenza A viruses (IAV) [5].

Viruses are great model organisms for evolutionary biology, not only because they have broad implications on global health and biogeochemical processes, but also for addressing fundamental questions. Viruses have relatively simple and compact genomes due to the constraint of being dependent on cellular organisms for replication. Thus, their genomes are the simplest system to model complex genome-wide interactions such as epistasis and clonal interference [6]. Viruses also are the smallest microbes that are at the boundary of living and non-living; thus they provide a key platform to understanding the central biological questions such as the origin of life, recombination, selfish elements, and host-pathogen coevolution [1], [7], [8].

Recent advances in data production and data analysis relating to the field of genetics open an opportunity to investigate these questions using big genomic datasets and novel computational methods [9]. In this chapter, the potential of viruses as a model organism is explored from the perspective of experimental evolution to that of global virology.

## 1.1 Experimental evolution of viruses

Viruses evolve in a complex cycle of transmission between hosts; thus, it is hard to predict the evolutionary trajectory of these pathogens in nature. To better understand the evolution of pathogens on a population-scale (or even on a global-scale), experimental evolution procedures have been developed as the simplified version of the complex world. Since these procedures control the environment during evolution, the number of necessary assumptions is minimized [10]–[12]. Furthermore, since the genome evolution of these pathogens can be observed temporally, more biological factors can be integrated into evolutionary models. For example, critical evolutionary processes such as drift, epistasis, clonal interference, and large offspring variance can be considered, factors that have mainly been ignored under large-scale pandemic models. These experimental studies may elucidate how pathogens evolve at larger and more complex scales in nature.

The standard procedure for experimental evolution of pathogens is to serially passage the microorganism in cell cultures under the presence or absence of a given treatment (i.e., drug/disinfectant agent). Samples of the pathogen are then collected and sequenced in a time-serial manner, and temporal allele trajectories of all nucleotides are generated. Due to the small size of virus genomes, population-level whole-genome sequences are accessible at low cost. These datasets are analyzed with the goal of estimating population genetic parameters such as effective population size ($N_e$) and selection coefficients ($s$). These parameters are essential in experimental evolution, as temporal changes in allele frequency can be deterministic due to fitness advantages (selection), or stochastic due to finite population sizes (genetic drift). Other model parameters such as mutation rate and recombination rate are important for understanding these processes that increase genetic variations enable viruses to adapt to new environments.

## 1.2 Model parameters in virus genome evolution

In virus genome evolution, a number of efforts have been made recently to infer parameters such as effective population size ($N_e$), selection coefficients ($s$) and mutation rate ($\mu$) from time-sampled experimental evolution data [10]–[13]. These methods are mostly based on the Wright-Fisher model (where population census size is assumed to be constant and discrete, utilizing random sampling of biallelic gene copies), with different approximations to infer population genetic parameters from temporal allele trajectories. These model parameters are the first steps to understanding the landscape of virus evolution, with the eventual goal of finding solutions to the problems of global viral epidemics. For example, the inference of effective population size ($N_e$), selection coefficients ($s$) and mutation rate ($\mu$) in an experimental evolution setting may provide the initial clues of the potential threat of viral pathogens adapting to a new drug or treatment in nature [10]–[13].

**Wright-Fisher model**
The Wright-Fisher model is one of the most fundamental models proposed by Sewall Wright and R.A. Fisher to represent the random process of allele frequency changes in finite populations [14], [15]. The model generalizes the genetic drift of alleles from one generation to the next as a binomial distribution, in the absence of other evolutionary processes such as selection, gene flow, and mutation. The binomial distribution defines the probability of sampling $k$ alleles in a sample of $N$:

$$\Pr(X = k) = \binom{N}{k} f^k (1-f)^{N-k} \quad (1)$$

where $f$ is the allele frequency. For viruses, $N$ represents the haploid population size. The assumptions made to simplify this process in virus population dynamics include discrete and non-overlapping generations, constant population sizes, and independent alleles. In the absence of other forces, the probability that an allele gets fixed in the population under genetic drift is equivalent to its initial allele frequency. The general consequences of the Wright-Fisher model in modeling genetic drift in finite populations are: the change in allele frequency due to genetic drift is random, the magnitude of genetic drift is bigger in smaller populations, and an equilibrium state of a population under the influence of only genetic drift is fixation or loss of an allele.

**Effective population size**
The census population size $N$ is the number of individuals in a population – however, not all individuals contribute to the changes in allele frequency in most biological populations. Thus, the genetic size of a population is defined as the effective population size $N_e$, which is proportional to the magnitude of genetic drift. $N_e$ can be determined by comparing the rate of genetic drift in a census population with the rate of genetic drift in an idealized population under the assumptions of the Wright-Fisher model.

There are two ways to estimate the effective population size of a sample population: Inbreeding effective population size ($Ne^i$) and Variance effective population size ($Ne^v$). The estimates from these two definitions of effective population sizes may differ when model assumptions for each definition are not met. For viruses, only the concept of the variance effective population size is relevant as viruses are haploid populations. The variance effective population size, $Ne^v$, can be estimated through expressing the changes of allele frequency over time in many replicates as a variance:

$$Var(\Delta f) = \frac{f_{t-1}(1-f)_{t-1}}{2Ne^v} \quad (2)$$

where $f$ is the allele frequency of an allele of interest, and $t$ is the number of generations.

When the population size fluctuates over time, as it does occur in viruses inter-host population bottlenecks and intra-host population growth during infection, the effective population size can be calculated using the harmonic mean:

$$\frac{1}{Ne} = \frac{1}{t}\left[\frac{1}{Ne_1} + \frac{1}{Ne_2} + \cdots + \frac{1}{Ne_t}\right] \quad (3)$$

where $t$ is the total number of generations over the fluctuating sizes. This harmonic mean gives more weight to small values due to the inverse of population size – thus, the genetic bottleneck is an important factor in determining the strength of genetic drift in a population over time.

**Selection coefficient**

Natural selection occurs when one phenotype causes a greater chance of survival and reproduction and the genotype that causes this phenotype increases in frequency over generations. The fitness of an individual can be defined by absolute fitness or relative fitness. Assuming two genotypes A and B, the absolute fitness of A is given as:

$$Absolute\ fitness\ of\ A = \frac{N_A}{N_A + N_B} \quad (4)$$

where $N_A + N_B$ is the total population size. Alternatively, the ratio of genotype-specific growth rates of A to B is given as the relative fitness:

$$\frac{N_B(t)}{N_A(t)} = w^t \frac{N_B(0)}{N_A(0)} \quad (5)$$

where $w$ is the strength of natural selection and $t$ is the number of generations. When $w > 1$, the genotype B is growing faster than the genotype A, and when $w < 1$, the genotype A is growing faster than the genotype B. Thus, the change in the frequency of the genotype A due to natural selection is given as:

$$\Delta f_A = \frac{f_t w_A}{f_t w_A + (1-f_t)w_B} - f_t \quad (6)$$

$$s_A = 1 - w_A \quad (7)$$

where $f_t$ is the frequency of the genotype A at the generation $t$. As shown above, fitness can be expressed in terms of selection coefficient, $s_A$, by taking the difference between the relative fitness and one. As viruses are entirely dependent on host cells for replication, they experience strong selective pressures during viral life cycles due to host immunity and/or drug treatment, both in experimental evolution and in nature.

**Mutation rate**
Since mutation is the ultimate source of genetic variation for all organisms, there has been much effort to understand the effect of mutation rate on the course of evolution. For host-dependent viruses, mutation is a vital mechanism for switching between hosts and evading host immune system. Particularly in experimental evolution, it is common to start serial passaging of viruses with a clonal population [10]–[12] – thus, the high mutation rate of viruses is responsible for creating the diversity of a virus population over the course of evolution. To calculate the mutation rate, the pairwise nucleotide diversity of an idealized population is assumed to be directly proportional to $\mu N_e$, where $N_e$ is the effective population size, and $\mu$ is the mutation rate, according to the neutral theory of molecular evolution [16].

Recent evidence at whole-genome and single-gene levels shows that mutation rates tend to evolve by selection to improve replication fidelity, whose variation is in turn created under random genetic drift [17]. This theoretical work on the mutation-rate evolution is based on the fact that most mutations are deleterious [16], which is especially true for viruses whose genome is compact and dedicated only to protein-coding. A mutator that increases mutation rate (such as a DNA polymerase variant, or a DNA repair protein) is subjected to the opposing forces of mutation pressure (input) and selection/recombination (output) [17]. The degree to which selection can reduce the mutation rate is dependent on the effective population size ($N_e$) that determines the strength of genetic drift. This drift-barrier hypothesis postulates that when the $N_e$ is small – thus, with strong genetic drift – the efficiency of selection in removing mutators decreases. For example, when viruses are subjected to mutagenic agents that increase mutation rate above the natural equilibrium, the response of virus genomes to such a rapid change in this vital mechanism is found to influence the effectiveness of these treatments against viral pathogens [12].

**Recombination rate**
Recombination is the exchange of genetic materials between chromosomes, and sexual reproduction is the production of new organisms by combining genetic information from two parents through recombination [18]. Recombination occurs in diploid eukaryotic organisms during reproduction, and also in prokaryotic cells and viruses when genetic material is transferred from a donor to a recipient. Recombination generates genetic variation and is also involved in other functions such as DNA repair. The ubiquity of recombination and sexual reproduction across the tree of life is one of the biggest questions in evolutionary biology. Evolutionary explanations for the advantages of recombination argue that it generates greater variability by breaking down genetic associations [19].

Viruses undergo genetic change by antigenic drift (point mutation) and antigenic shift (recombination or re-assortment). Genetic recombination is frequent in both RNA and DNA viruses and occurs at highly variable rates. For example, the rate of recombination per nucleotide in Retroviruses (ssRNA-RT) like HIV exceeds that of mutation, though it can vary from high to nonexistent in other RNA viruses [7]. Similarly, segmented viruses exhibit re-assortment rates ranging from low (e.g., Hantaviruses) to high (e.g., influenza A virus) [7].

Recombination/re-assortment allows viral genomes to undergo major genetic changes during evolution, such as increasing virulence and pathogenesis, evading host immunity, altering transmission tactics, expanding host ranges, evolving antiviral resistance, and potentially creating new viruses. There is an ongoing debate on whether recombination is a form of sexual reproduction for viruses, as is for cellular organisms [7]. Despite the extensive results that the benefits of recombination are responsible for the ubiquitous presence of sexual reproduction [19], the viral community asserts that there is little evidence of recombination being favored by natural selection for viruses and that it is a mechanistic by-product of other aspects of viral biology [7].

Recombination is a particularly important event in the viral epidemic since it allows viral genomes to undergo major genetic changes. Furthermore, viral genomes are almost entirely functional which makes most mutations deleterious. Recombination is therefore essential for breaking down negative genetic associations in these cases, as any advantageous allele is likely to be linked with other deleterious mutations. Other stochastic events such as genetic drift (due to small effective population sizes within hosts) and sweepstake events (due to stochastic variation in reproductive success) during host-to-host transmission are significant factors impacting the prediction of viral evolution.

## 1.3  From experimental evolution to global virology

We can build on the knowledge acquired from experimental evolution to study the complex and extensive network of viruses in nature, with the goal of predicting the evolutionary trajectories of viruses on a global scale. Furthermore, the population genetic knowledge acquired from the experimental evolution studies may help investigate the role of viruses in the long-term evolution of various ecological systems, such as microbial communities and human microbiota, given the ubiquitous presence of viruses. Investigating viral biology on a global scale is important for the following reasons.

Firstly, viruses have the potential to become a threat to global public health as shown by the recent outbreaks of Ebola virus[1] and Zika virus[2], aided by the ever-increasing mobilization of human populations. Moreover, the possibility of air-borne and water-borne transmission shows how these pathogens can transmit rapidly between their hosts. The emergence of new viruses may occur unpredictably by forming new subtypes through antigenic drift and/or antigenic shift. In viral epidemiology, the recent advances in genetic data production and analysis are helping to predict the evolutionary path of viral pathogens, based on the previous data and intricate models [20], [21].

Secondly, viruses play vital roles in the biochemical processes of global microbiota. It is estimated that marine viruses are responsible for over one-third of host mortality per day through lysing host cells, and over 1,000 gene movements per day through horizontal gene transfer [22]. In cold environments, the impact of virulence and viral-host evolution on soil biogeochemical cycling is limited – however, the conditions in cold soils such as permafrost and glaciers are changing due to accelerated climate change. The impact of viruses continues to increase on the ecosystem of microbiota in climate-sensitive areas. Thus, investigating viral abundance and diversity to characterize the Earth's virome through metagenomic data is vital [23].

---

[1] Ebola virus disease. (2019.02.01). Retrieved from https://www.who.int/news-room/fact-sheets/detail/ebola-virus-disease

[2] Zika virus. (2019.02.01). Retrieved from https://www.who.int/news-room/fact-sheets/detail/zika-virus

Thirdly, viruses have vast implications in the evolution of the Earth's biomes – the role of viruses is essential is evolving and regulating microbial communities, and the extent of its influence is ongoing research. Phages are viruses that infect microbes, and virus-host interactions result in coevolution that often leads to biological novelties. Phages are selfish replicators that propagate their genomes through infecting microbes – thus, microbes evolve defense mechanisms such as CRISPR/Cas adaptive immune system against phage infections, and phages co-evolve to evade these mechanisms [24]. Viruses have two distinct states of reproduction: virulent (lytic) and temperate (lysogenic) [25]. Lytic phages affect the mortality of microbes, and this is the major force behind the organic matter cycling in various ecosystems. Lysogenic phages are integrated into the chromosome of microbes, and these prophages replicate together with their hosts. As the fitness of prophages is directly linked to their host, new dynamics between hosts and pathogens arise, such as inhibiting superinfection of the host by other phages and providing virulence factors to their hosts to invade new niches. Lysogeny can also gradually enable phages to evolve from lytic phages (complete selfishness) to cryptic phages (complete domestication) where they become unable to form infective particles and evolve as one selectable unit with the host.

## 1.4 Likelihood-free inference in virus genome evolution

Advances in genomics allow the scientific community to move from small-scale inference such as experimental evolution to large-scale inference such as viral epidemics. In conventional evolutionary models, statistical inference with analytic approaches has been the focus of data analysis. However, it may be difficult to compute likelihood functions from models with many latent variables and/or complex architecture [26]. This problem is particularly prevalent in biological modeling since biological systems in nature tend to be very complex– such as phylogenetic trees, gene networks, and ecology systems to name a few. Thus, numerous likelihood-free methods have been developed to overcome this limitation in model-based inference, including Approximate Bayesian Computation, where a likelihood function is replaced by simulation [27]–[29], and Variational Inference, where intractable integrals are approximated using standard probability distribution [30], [31]. This chapter discusses some of the popular simulation-based methods used to infer the parameters of a statistical model, with a particular focus on their application in virology.

### 1.4.1 Approximate Bayesian Computation

One of the best ways to describe observed data is by fitting a statistical model with known or/and unknown parameters. In Bayesian frameworks, the inference of model parameters from data is carried out by computing the posterior distribution $P(\theta|Data)$ from the prior distribution $P(\theta)$ and likelihood function $P(Data|\theta)$ as follows:

$$P(\theta|Data) = \frac{P(Data|\theta)P(\theta)}{P(Data)} \quad (8).$$

However, when the model is very complex, as in population genetics, the exact likelihood calculations can become intractable due to the presence of latent variables or heavy computation [26]. Approximate Bayesian Computation (ABC) is used to overcome this problem [32], [33]. ABC replaces likelihood calculations by simulating the model of interest directly with random input values from its prior distribution (Figure 1). This allows very complex models to be inferred since simulating them with a set of input values is much easier and faster than calculating the likelihood of each parameter space [32]. These simulations are classified using summary statistics, which are then compared to observed data to construct a posterior distribution for the parameter of interest (Figure 1).

In ABC, summary statistics are used to summarize high-dimensional data such as a sample of multivariate measurements into low-dimensional statistics comparable to the observed values. However, as the number of summary statistics increases to describe complex data more accurately, a "curse of dimensionality" occurs in which most simulations get rejected due to increasing dimensionality. For example, if three summary statistics are accepted with 10% tolerance, 99.9% of the simulations will be rejected due to the three-dimensional tolerance space [26]. Furthermore, an unspecified loss of information by using insufficient summary statistics may lead to the problem of choosing the wrong model with ABC [34].

ABC model choice is used to choose between two models: $M_0$ as a null model, and $M_1$ as an alternative model. The relative probability of $M_1$ over $M_0$ can be computed through the model posterior ratio as the Bayes factor $B_{1,0}$ [33]:

$$\frac{p(M_1|D)}{p(M_0|D)} = \frac{p(D|M_1)p(M_1)}{p(D|M_0)p(M_0)} = B_{1,0}\frac{p(M_1)}{p(M_0)} \quad (9)$$

when the model prior $p(M_0)$ is equal to $p(M_1)$. In practice, the model priors are made equal by producing the same number of simulations for each model and retaining the best simulations from the combined simulations. In ABC model choice, the posterior ratio is computed as the number of accepted simulations from $M_1$ over those of $M_0$ – giving the Bayes factor $B_{1,0}$ which is an indicator of the support for a specific model.

As ABC has become an important tool for inferring parameters with complex models, many efforts are being made to improve the method [33], including integrations of Markov Chain Monte Carlo (MCMC) [29] and Hamiltonian Monte Carlo (HMC) [35]. These methods increase the power of ABC inference by improving the random sampling of input values from the prior distribution of a parameter. Markov Chain Monte Carlo (MCMC) is a widely used algorithm for sampling and attempts to solve the problem of sampling from a complex distribution. Biological models are particularly complex, thus sampling from a complex distribution used to represent these models leads to the problem of the probability of drawing a desired outcome very low. Thus, MCMC is introduced as a randomized algorithm that makes the sampling process more efficient by making the probability of sampling an output approximately equal to the target probability distribution. There are two steps in this algorithm: 1. Monte Carlo step which is a random walk taking a large sample, 2. Markov Chain step which estimates the expectation of the probability distribution. Examples of Monte Carlo methods include Metropolis-Hastings algorithm that uses a proposal density for the next move that can be accepted or rejected and Gibbs sampling that uses the conditional distributions of the target probability distribution. These methods are particularly efficient when sampling from high-dimensional distributions.

### 1.4.2 Simulation-based inference in virology

Simulation-based methods such as Approximate Bayesian Computation (ABC) are particularly useful in population genetics, where models to represent the forces of evolution (genetic drift, selection, gene flow, mutation) at a population-level become complex and dynamic. For this reason, the applications of ABC have been pioneered and widespread in population genetics – to name a few, for inferring growth rates and time of divergence from genetic data, and for model choice between competing models of human demographic history [26]. Other examples of ABC in the fields of evolution and ecology include phylogeography, systems biology and epidemiology [26].

In virology, simulation-based methods have been successfully applied to infer population genetic parameters such as effective population sizes and selection coefficients from the Wright-Fisher models. For example, according to the simulation studies comparing the performance of different time-serial methods for inferring population genetic parameters, the simulation-based method is shown to perform the best for virus populations that experience high selective pressures due to their dependency on host cells [36].

Despite their practicality, the performance of simulation-based methods is limited to the particular model used for inference. For example, the Wright-Fisher model used in these inference studies assumes mutations arising in the course of evolution of viruses are completely independent. This assumption is not valid even in the simplest organisms like viruses, especially because their genomes are entirely functional, making potential interactions between mutations prevalent through compensation (epistasis) or competition (clonal interference). Thus, the conventional model-based inference, including simulation-based methods, has limited abilities to learn new features biological from data.

## 2 Artificial Neural Networks

Artificial Neural Networks (ANN) is a model-free approach that has a great potential for applications to big genetic data. Deep learning using Artificial Neural Networks is a subfield of machine learning, where computers are able to extract patterns from raw data and acquire their own knowledge of the real world. This task is done by representing data as a piece of information known as a feature – for example, the features about a patient such as the previous medical records enable machine learning algorithms to predict the diagnosis of the patient. However, it is often difficult to manually extract the best features that might be useful for machine learning algorithms. This downside is particularly problematic in complex tasks such as recognizing objects in pictures since these objects such as cars or dogs can take various shapes and colors. One solution is to use a machine learning technique called representation learning to learn features themselves instead of only mapping representation to output [37]. However, extracting high-level features from raw data is not an easy task. Deep learning tackles this problem by expressing representations in terms of simpler representations in a hierarchical manner. This advantage provides a flexible framework where features are easy to adapt and learn, without manual over-specification of features like other machine learning techniques. Historically, the fundamental ideas and architectures of deep learning have been inspired by biological neural networks in the brain; hence it goes by the name of Artificial Neural Networks or Multilayer Perceptron (MLP) or Feedforward Deep Network. In this chapter, the basic concepts of artificial neural networks are introduced, leading to the discussion of their potential applications in virology.

### 2.1 Feedforward Neural Networks

The goal of neural networks is to approximate a function $f^*$. A feedforward network defines a mapping $y = f(x; \theta)$ and learns the parameters $\theta$ that result in the best function approximation. In this feedforward, the information flows from the input $x$ through the intermediate computations defining $f$ to the output $y$, with no feedback (unlike recurrent neural networks). In mathematical terms, a multilayer perceptron is a function mapping input to output, which is composed of simpler functions. For instance, a multiplayer perceptron may be composed of a chain of three functions $f^{(1)}, f^{(2)}$, and $f^{(3)}$;

$$f(x) = f^{(3)}(f^{(2)}\left(f^{(1)}(x)\right)) \quad (10)$$

where $x$ is the input, $f^{(1)}$ is the first hidden layer, $f^{(2)}$ is the second hidden layer, and $f^{(3)}$ is the output layer of the networks as shown in Figure 2. The length of the chain is called the depth of the network, and it is where the concept of "deep learning" arises.

Within a neuron, an elementary unit shown as blue circles in Figure 2, the input values in each hidden layer are multiplied by their corresponding weights (*w*) and added by a constant bias term (*b*):

$$f(x; w, b) = x^T w + b \quad (11)$$

where the function used in hidden layers is called the activation function. Weights control the strength of connection and bias offsets the output while independent of data. Depending on the neural networks, the activation function can take diverse non-linear forms such as sigmoid $f(x) = 1/(1 + e^{-x})$ or ReLU $f(x) = \max(0, x)$ functions. During training, the data provide noisy and approximate examples $x$ of $f^*(x)$ to drive $f(x)$ to match $f^*(x)$. The output layer must produce a value close to $y$ without being told what intermediate layers should do (thus called hidden layers) – the learning algorithm implements an approximation of $f^*$.

## 2.2 Training Neural Networks

To train neural networks, we need to do the following steps:
1. Define a loss function (=cost/objective/error function) that quantifies the inconsistency between predicted value $\hat{y}$ and actual label $y$.
2. Find the parameters that minimize the loss function (optimization)

### 2.2.1 Loss function

Given the dataset $\{(x_i, y_i)\}_{i=1}^{N}$, where $x_i$ is the input and $y_i$ is the output/label, the loss over the dataset is a sum of loss over the examples:

$$L = \frac{1}{N} \sum_i L_i(f(x_i, W), y_i) \quad (12)$$

Furthermore, the loss function has a regularization term (for example, L2 regularization tends to prefer smaller and spread out weights $W$) that imposes a penalty on the complexity of a model to prevent overfitting.

$$R(W) = \sum_k \sum_l W_{k,l}^2 \quad (13)$$

Thus, the loss function is given as:

$$L(W) = \frac{1}{N} \sum_{i=1}^{N} L_i(f(x_i, W), y_i) + \lambda R(W) \quad (14)$$

where $\lambda$ is the regularization strength, which is a hyperparameter to be chosen.

### 2.2.2 Optimization

In order to find the values of W that reduce a loss function, we can use the gradient descent:

$$\frac{df(x)}{dx} = \lim_{h \to 0} \frac{f(x + h) - f(x)}{h} \quad (15)$$

The computation of an analytic gradient is fast and exact but error prone. The computation of a numerical gradient is easy to write, but slow and approximate. To overcome the challenges of computing gradients over a huge neural network, the backpropagation algorithm computes the gradient cheaply by allowing the information from the cost to flow backward through the network. It works by splitting the gradient into local gradients with the chain rule to iteratively compute gradients for each layer.

## 2.3 Artificial Intelligence in virology

The concept of Artificial Intelligence (AI) arises from the futuristic vision that machines can become intelligent by having the ability to learn and reason like humans. Two aspects of modern society are driving AI to be a thriving field - computational power and big data. Computational biologists also envision programmable computers that can mimic human intelligence to learn from biological data in the near future. The application of machine learning methods in biological systems allows model-free investigations, particularly through the rapid progress in deep learning, where neural networks are used to learn functional relationships from observed data [38]. Deep learning has the potential to solve complex problems such as predicting viral epidemics using big genomic datasets, through multilayer neural networks, which were previously thought infeasible [38].

Statistical models in viral epidemics have to consider multiple factors and make strong assumptions, due to the complexity of viral biology, host biology, and host-pathogen interactions. As viruses only replicate within their hosts, they tend to evolve either a symbiotic relationship with their hosts that may be antagonistic (pathogenic), commensal (hitchhiking), and mutualistic (beneficial) [39]. This complexity in the viral life cycle makes the prediction of viral epidemics difficult, as predictive models have to incorporate events that occur at the micro-level within hosts and the macro-level between hosts. For instance, these models have to consider viral biology that is highly diverse between different virus types such as in virulence, replication, and recombination at the micro-level [1]. Furthermore, there are population genetic factors such as effective population size, mutation rate, and host defense systems that affect the evolutionary trajectories of viruses even within the same virus type. Additionally, these models have to incorporate factors that affect the transmission between hosts at the macro-level, such as migration patterns and transmission routes of their hosts.

Despite recent efforts, model-based methods have difficulties in incorporating these factors due to the complexity of solving or simulating these models [40]. Furthermore, these model-based methods do not take full advantage of hidden patterns in high dimensional biological data and/or big genetic data. To solve intricate problems such as viral epidemics, machine learning methods are promising tools to drive model-free and data-driven research to mine hidden patterns and make accurate predictions. Artificial neural networks are particularly useful for such problems, as they can extract features from data in a hierarchical manner [37].

We now have access to big genomic datasets that are essential resources for machine learning approaches. For example, Influenza Research Database (fludb.org) currently has 661,000 segment sequences available[3], which offers a great opportunity, but also an overwhelming challenge for traditional model-based approaches. Indeed, the scientific bottleneck no longer occurs at the level of genomic data production, but rather at the level of data analysis. Moreover, other big data such as digital epidemiology and global human migration patterns are growing exponentially. Using artificial neural networks and genomic sequences from the Influenza Research Database, a predictive platform may be generated at the micro-level and/or micro-level. Using influenza virus (IAV) genome sequences and other epidemiologic data from the previous years as training examples, an algorithm may be designed to predict IAV strains likely to be prevalent in the following year. The current challenge is to curate different types of data as input and to design the architecture that can integrate both the events that happen at the micro-level within hosts and at the macro-level between hosts.

---

[3] Influenza Research Database. (2019.01.30). Retrieved from https://www.fludb.org/brc/dataSummary.spg?decorator=influenza

# 3 Third-Generation DNA sequencing

In the field of biology, advances in technology and engineering are expanding the scope of questions that can be investigated with theoretical and computational tools. For example, DNA sequencing is becoming more efficient, accurate, and cost-effective each year, such that a human genome can be sequenced for only $1,000 at 30x coverage, and at a speed of 18,000 genomes per year[4]. This innovation is phenomenal considering that the first full sequencing of the human genome was only completed in 2003. These rapid improvements in DNA sequencing allow more samples to be collected, from different individuals and/or at different time points.

Since the emergence of the Second Generation Sequencing in the late 1990s, lowered costs and improved efficacy of DNA sequencing is enabling the production of multiple time-point datasets [41], which are particularly revealing for the studies conducted in the domains of ancient DNA [41], experimental evolution [36], and clinical trials [42]. These datasets allow temporal observation of rapidly evolving organisms at different time-points in the past [10]–[12]. The studies of virus genome evolution benefit from the information on the temporal aspects of evolutionary forces, such as changes in selection, population sizes, or mutation rates [10], [12], [13]. Moreover, we can gain insights into the complex aspects of evolutionary trajectories, such as host-pathogen interactions or intra-/inter-species competition and cooperation.

Big genetic data generated from multiple individuals provide the means to investigate the genetic variants of a population, as shown by the success of genome-wide association studies (GWAS) [43]. For viruses, sampling multiple individuals at the whole-genome level is feasible due to the small genome size of viruses. However, most virus datasets have been produced with the technique of pooled sequencing, where genetic materials from several individuals are pooled for sequencing. While this method of pooled sequencing is cost-effective and time-efficient, individual haplotypes cannot be reconstructed as in individual sequencing [12]. The novel technologies of long-read sequencing are pioneering platforms to generate the genetic data from multiple individuals cost- and time-effectively, including viral pathogens. In this chapter, the recent advances in sequencing technologies and their impact on virus research are introduced.

## 3.1 DNA sequencing technologies

**First-Generation Sequencing (1970)**
The structure of DNA was first modeled by James Watson and Francis Crick in 1953 [44], based on X-ray results by Rosalind Franklin. The first method of DNA sequencing was developed in the 1970s [45], [46], and the method developed by Frederick Sanger et al. in 1977 was introduced as one of the first commercially available techniques known as Sanger sequencing [46]. This method uses the selective incorporation of chain-terminating dideoxynucleotide (ddNTPs: ddATP, ddGTP, ddCTP, ddTTP) by DNA polymerase during *in vitro* DNA replication. Sanger sequencing was the most widely used sequencing method for the next 40 years.

---

[4] HiSeq X™ Series of Sequencing Systems. (2016.02.10). Retrieved from http://www.illumina.com/content/dam/illumina-marketing/documents/products/datasheets/datasheet-hiseq-x-ten.pdf

**Second-Generation Sequencing (2000)**

By the year 2000, several new platforms were developed to achieve high-throughput sequencing method for DNA sequences. One of the most commercially successful products is from Illumina's Hi-Seq sequencers, which uses clonal amplification and sequencing by synthesis to allow parallel sequencing. The development of second-generation sequencing enabled more efficient and accurate DNA sequencing for a lower cost, for example, lowering the cost of sequencing a human genome from US$1 million down to US$4,000 by 2013. Additionally, the rapid data production of DNA sequences has pushed the scientific community to develop new computational methods for sequence processing, storage, and analysis.

**Third-Generation Sequencing (2015)**

Most recently, another innovation in DNA sequencing has emerged using new methods that allow long-read sequences, as compared to the previous methods. The most notable techniques are Single-Molecule-Real-Time (SMRT) sequencing and Nanopore sequencing. SMRT sequencing is based on sequencing by synthesis, where DNA sequences are synthesized in small containers at the bottom of the well, and fluorescently labeled nucleotides are incorporated by DNA polymerase to be detected by the smallest light detector volume in the well. Nanopore sequencing detects the changes in ion current of a nanopore when a DNA sequence passes through it. Both technologies are revolutionizing DNA sequencing by the production of long-read DNA sequences (>10,000 nucleotides as compared to typically 100 nucleotides in the Second Generation Sequencing) in real-time.

## 3.2 Long-read sequencing for virology

The Third-Generation DNA sequencing has emerged in the last few years using new technologies that allow the generation of substantially longer read sequences. In particular, the portable Nanopore sequencing device, MinION, can be used for de novo sequencing and metagenomics, and they are being tested for high-impact environmental research and pathogen analysis [47], [48].

Only a few years after the successful launch of the long-read sequencers, the scientific community is experiencing a rapid change in how genetic data are generated at the frontier research topics. For instance, since the first paper appeared in 2013 with the keyword "Nanopore" at the preprint server for biology[5], the number of papers with the same keyword has been doubling every year until 2018 as shown in Table 1. At this trend, genetic studies with the Third-Generation sequencing are expected to become the norm of genetic research studies in a few years.

In virology, such innovations are also pervasive – various viruses are being sequenced with the long-read sequencer, including Ebola virus, chikungunya virus, dengue virus, yellow fever virus, and influenza A virus to name a few [48]–[52]. Since virus genomes are small and compact, long-read DNA/RNA sequences (>10,000 nucleotides) cover the complete genomes of many viral pathogens. Viral identification is made more efficient and accurate through such sequences in metagenomic samples. Other than reading long-read sequences, the Third-Generation sequencers such as Nanopore have further advantages for virus sampling. The portability and cost-effectiveness of these sequencers are enabling rapid and field-based analysis of pathogens on-site at resource-limited settings. For an epidemic response, early identification of the viral strain of a new outbreak is crucial in monitoring and containing the epidemic through effective vaccination [48], [52]–[54]. In the clinical setting, the capacity to sequence in real-time offers a revolutionary diagnostic tool for various viruses that can cause disease outbreaks in human [49], [51] and livestock [55], [56].

---

[5] BioRxiv: the preprint server for biology. (2019.02.07). Retrieved from https://www.biorxiv.org/

Furthermore, the ability to sequence DNA/RNA directly facilitates the characterization of genetic plasticity by eliminating the PCR step, for investigating complex topics such as epigenetic modification and transcriptome profiling [57]–[59]. For instance, long-read sequencing has greater efficiency in sequencing full-length transcripts and identifying RNA isoforms [59]. In metagenomics, these advantages of highly parallel direct RNA sequencing also improve the assembly of viral genomes and the investigation of microdiversity [60]. As long-read sequencing is revolutionizing the quantity and quality of genetic data, an innovation in the downstream analysis using automated machine learning methods is an urgent task to achieve in computational biology.

# 4 Futuristic methods in virus genome evolution

The scientific community is experiencing a rapid change in the research paradigms, both through the advances in data production and data science. An ever-increasing growth of big data is evident in a wide variety of research fields, from computer vision, astronomy, weather forecasting, health care, to genetics. To process such data, researchers are pushing the boundaries of computational power and method. Novel computational approaches such as machine learning methods are being developed to learn from big datasets, and these advances in data science are inspiring researchers from other fields to apply those methods in their respective datasets. In computational biology, researchers attempt to apply these novel methods using machine learning to process and classify high dimensional biological data, to predict future events or to acquire new knowledge.

Model-free approaches such as machine learning methods eliminate the errors resulting from strong assumptions about underlying mechanisms by taking a data-driven approach [40]. The digitization of society has opened the age of big data, making machine learning methods more attractive by lightening the key burden of statistical estimation from small datasets [37]. This data-driven research takes full advantage of the advances in data production by learning directly from the vast amount of data produced (unsupervised learning). Furthermore, model-free methods may change the focus of study from inference to prediction, as machine learning methods can be designed to learn from previous input-output pairs to predict the output from a new sample (supervised learning). Unlike inference, the performance of predictive algorithms may be assessed directly with future data.

Advances in model-free approaches are particularly important for virus evolution. We aim to predict the course of viral pathogens efficiently and accurately from previous data, which is vital for critical societal and economic issues such as global public health and agricultural industry (supervised learning). Furthermore, we aim to acquire new knowledge on the ecology and evolution of viruses in various systems, through big genomic data of new viruses (unsupervised learning). In this chapter, the future of virology is discussed in conjunction with advancing sequencing technologies and data analyses, using specific examples in supervised learning and unsupervised learning setting.

## 4.1 Futuristic big data analysis for virology

The rapid increase in genetic data leading to the current challenge of dealing with big data is expected to accelerate with the Third-Generation sequencing technologies. Applications of long-read sequencing enable real-time data production without an intermediate data processing step, as shown in Figure 3. Until recently, considerable efforts have been devoted to the upstream data processing of the Second-Generation sequences - for assembly, alignment and quality control of short sequence contigs. In futuristic approaches, the conventional workflow of the Second-Generation Sequencing may be simplified into real-time data production and on-site data analysis using the Third-Generation Sequencing and machine learning analysis, respectively. This automatization is an important improvement to the diverse fields of virology such as experimental evolution as well as environmental and medical sampling campaigns. Long sequences from the Third-Generation sequencers eliminate the errors and uncertainties resulting from assembling short sequence contigs into a physical genome map in the Second-Generation sequencing. Thus, an innovation in the downstream data analysis step is necessary to analyze large-scale data generated from real-time long-read sequencing (Figure 3).

It is of a future challenge in computational biology to achieve automated machine learning approaches that minimize human input to analyze real-time data from long-read sequencers on-site. Machine learning combines pattern recognition and computational learning to perform predictive data analysis. As shown in Figure 3, the application of neural networks innovates the downstream data analysis to process genetic data from the Third-Generation sequencing by big data analysis without model assumptions or feature specifications, as required in the classical bioinformatics.

## 4.2  Supervised Learning: virus classification

Supervised learning aims to predict an output from an input, based on previous input-output pairs. It infers a relation that maps input data to output data by learning a function that maps them through labeled training examples. Thus, supervised learning requires a big set of labeled training examples and a great deal of effort is put into gathering and labeling a training set from human experts or automated measurements. For artificial neural networks, the step to represent raw data as features is not necessary, but the architecture of neural networks (i.e., the number of neurons and the depth of layers) is a major factor that influences the performance of supervised learning. Then, the learning algorithm can be run on the labeled training set to increase the accuracy of the learned function for an optimized prediction. The performance of the learned function is measured through a test set that is unseen during the training set. As mentioned before, the advantages of artificial neural networks in learning non-linear and complex functions over several layers of neurons [37] apply to high dimensional genetic data.

In virology, the potential of supervised learning using artificial neural networks is significant. Viruses display a vast diversity in form and function, and there are already virus genome datasets available, which are essential resources to design a comprehensive platform for virus classification using supervised learning. Viruses can have a vast functional diversity comparable to the corresponding microbial populations [61]. Applications of new sequencing technologies and advanced computational methods can improve the current knowledge of viruses by providing relevant data and techniques to test hypotheses.

Despite their abundance, diversity and ecological significance in the biosphere, viruses are understudied in many ecological communities [62]. For example, the lack of knowledge on the impact of viruses is particularly relevant in rapidly changing ecosystems such as glaciers and permafrost [63]–[65]. Permafrost composes of approximately 25% of Earth's land surface and stores almost 50% of global soil carbon in a frozen form. Near-surface permafrost across the Arctic region is rapidly thawing due to climate change [66]. Recent studies reveal that permafrost soils have an extensive presence and diversity of viruses – however, their impact on the microbial populations in this ecosystem of microorganisms is yet to be understood [67], [68]. On the other hand, glaciers cover approximately 10% of Earth's land surface and host diverse microbial communities [64]. Along with surprising microbial diversity, viruses are also actively present in these cold ecosystems, with the rate of virus infection in these bacteria being one of the highest observed [65], [69]. Viruses in these permafrost- or glacier-covered soils experience strong selection pressures due to the extreme nature and isolation of these environments. Moreover, resident microorganisms have other key roles in global biogeochemical processes such as methane cycling, fermentation, degradation of complex molecules and carbon gas emissions. However, the difference in virus diversity and activity in these two rapidly changing ecosystems is yet to be studied and assessed. Given the ever-increasing amount of sequence data generated through high-throughput technologies, large-scale comparative analysis with metagenomics is an invaluable tool to investigate viral diversity and activity from such environmental samples. As an example of the futuristic approach to investigate the viral impact on the ecosystem of cold soils, field sampling campaigns in permafrost regions using novel portable and real-time devices for long-read sequencing may be used. After which, these samples may be analyzed on-site in real-time using machine learning methods for supervised learning. This will help us understand the virome in these climate sensitive regions and examine the impending impact of thawing on the global biogeochemical cycling and coupled viral-host evolution of microbial and viral communities.

Specifically, we can envision a portable processor chip for virus classification on-site of such sampling campaigns. In computer vision, one of the frontier fields in the application of neural networks, low-power and high-performance vision processor units including Movidius[6] are already commercially available. These are vision processor units that carry out inference of new examples from a trained network at very low power. As training takes most of the time and computer power, deep neural networks have already been trained with millions of labeled examples of images to learn highly non-linear and complex functions for image classification. Thus, these devices deploying deep learning can classify new examples quickly on-site, as inference/prediction using pre-trained neural networks is a computationally cheap task. For instance, Movidius may be applied for intelligent machine vision systems such as robotics and augmented & virtual reality to classify images as quickly as human vision is able to do. A processor chip with similar purposes is an example of futuristic methods in studying virus genome evolution – a device containing deep neural networks pre-trained with the vast amount of labeled virus genomes. These portable devices may be used along with long-read sequencing to provide an integrated pipeline of real-time data production and on-site data analysis, as shown in Figure 4. A researcher can bring a portable long-read sequencer and a portable processor chip of neural networks pre-trained with virus genomes to field sampling campaigns, from which diverse viruses can be sequenced and classified on-site. In the near future, such devices will revolutionize how sampling campaigns for clinical and environmental samples are undertaken, advancing the efficiency and pace of global virus studies for diagnostic purposes as well as for basic research.

### 4.3 Unsupervised learning: virus clustering

One of the biggest advantages of artificial neural networks is its inherent capacity to extract features from raw data in a hierarchical manner [37]. The ability to extract features without strong assumptions or manual over-specifications makes artificial neural networks suitable for unsupervised learning. Unsupervised learning mines hidden patterns, structures, or features from unlabeled test data through methods such as clustering and autoencoders. Such methods are useful at the exploratory stage of new studies without access to labeled datasets, to search for innovative questions and acquire pioneering knowledge.

---

[6] Movidius (2019.02.13). Retrieved from https://www.movidius.com/

In virology, unsupervised learning is a largely unexplored field despite its potential to solve challenges arising in the downstream analysis due to the increasing rate of data production. Recently, a new approach has been developed to test the possibility of applying Artificial Neural Networks (ANNs) to virus genetic data [6], based on the results from the simulation-based inference of population genetic parameters from an experimental evolution study [11]. In the study, a simple neural network was applied to the genetic datasets to execute unsupervised learning of virus genome evolution. This method was able to infer evolutionary distances from raw genetic datasets under the presence or absence of selective pressures in the experimental evolution setting. Figure 5 summarizes the workflow of unsupervised learning of virus genome evolution with the Nucleotide Skip-Gram Neural Network [6]. The first step is to train the Nucleotide Skip-Gram Neural Network using the deep learning software, where the genetic interactions between all significant mutations from the experimental evolution of virus [11] are learned as features of the neural networks. Subsequently, Principal Component Analysis (PCA) and hierarchical clustering are applied to produce a pairwise correlation map of these features, which reflects pair-wise evolutionary distances between all the mutations arising in the course of experimental evolution. Thus, this study demonstrates that mutations can be represented as distributed vectors that encode information of biological features and can be learned from data with artificial neural networks, rather than discrete entities within classical population genetic models, such as the Wright-Fisher model. This study shows that unsupervised learning with neural networks can achieve the same level of exploratory knowledge by clustering mutations of similar evolutionary trajectories automatically [6], as the human expertise had obtained from a series of several experimental studies [11].

In viral metagenomics, the features to learn are biological or evolutionary similarities that exist between billions of sequences in raw metagenomic data from different sources. Unsupervised learning from artificial neural networks has the potential to offer considerable benefits for futuristic data exploration: from field sampling of natural populations using the real-time sequencer to pattern mining of raw metagenomic data. This novel pipeline eliminates the intermediate step of data processing through the Third-Generation sequencing and the necessity of model assumptions through artificial neural networks. Therefore, these proposed futuristic methods have broad implications of improving field sampling and genetic data analysis, as well as of investigating the evolution and ecology of viral and microbial communities.

# 5 References

[1] E. V Koonin and V. V Dolja, "A virocentric perspective on the evolution of life," *Curr Opin Virol.*, vol. 3, no. 5, pp. 546–557, 2013.

[2] M. M. Tanaka and F. Valckenborgh, "Escaping an evolutionary lobster trap: Drug resistance and compensatory mutation in a fluctuating environment," *Evolution (N. Y).*, vol. 65, no. 5, pp. 1376–1387, 2011.

[3] A. R. Hall, P. D. Scanlan, A. D. Morgan, and A. Buckling, "Host-parasite coevolutionary arms races give way to fluctuating selection," *Ecol. Lett.*, vol. 14, no. 7, pp. 635–642, 2011.

[4] S. M. Andrews and S. Rowland-Jones, "Recent advances in understanding HIV evolution.," *F1000Research*, vol. 6, p. 597, 2017.

[5] E. J. Schrauwen and R. A. Fouchier, "Host adaptation and transmission of influenza A viruses in mammals," *Emerg. Microbes Infect.*, vol. 3, no. 1, pp. 1–10, Jan. 2014.

[6] H. Shim, "Feature Learning of Virus Genome Evolution With the Nucleotide Skip-Gram Neural Network," *Evol. Bioinforma.*, vol. 15, p. 117693431882107, Jan. 2019.

[7] E. Simon-Loriere and E. C. Holmes, "Why do RNA viruses recombine?," *Nat. Rev. Microbiol.*, vol. 9, pp. 617–626, 2011.

[8] E. V. Koonin and V. V. Dolja, "Virus World as an Evolutionary Network of Viruses and Capsidless Selfish Elements," *Microbiol. Mol. Biol. Rev.*, vol. 78, no. 2, pp. 278–303, 2014.

[9] E. C. Holmes, "Viral evolution in the genomic age.," *PLoS Biol.*, vol. 5, no. 10, p. e278, Oct. 2007.

[10] M. Foll, Y.-P. Poh, N. Renzette, A. Ferrer-Admetlla, C. Bank, H. Shim, A.-S. Malaspinas, G. Ewing, P. Liu, D. Wegmann, D. R. Caffrey, K. B. Zeldovich, D. N. Bolon, J. P. Wang, T. F. Kowalik, C. A. Schiffer, R. W. Finberg, and J. D. Jensen, "Influenza virus drug resistance: a time-sampled population genetics perspective.," *PLoS Genet.*, vol. 10, no. 2, p. e1004185, 2014.

[11] Q. Zhong, A. Carratalà, H. Shim, V. Bachmann, J. D. Jensen, and T. Kohn, "Resistance of echovirus 11 to ClO2 is associated with enhanced host receptor use, altered entry routes and high fitness," *Environ. Sci. Technol.*, vol. 51, no. 18, pp. 10746–10755, Sep. 2017.

[12] A. Carratala Ripolles, H. Shim, Q. Zhong, V. Bachmann, J. D. Jensen, and T. Kohn, "Experimental adaptation of human echovirus 11 to ultraviolet radiation leads to tolerance to disinfection and resistance to ribavirin," *Virus Evol.*, vol. 3, no. November, pp. 1–11, 2017.

[13] H. Shim, S. Laurent, S. Matuszewski, M. Foll, and J. D. Jensen, "Detecting and Quantifying Changing Selection Intensities from Time-Sampled Polymorphism Data.," *G3*, vol. 6, no. 4, pp. 893–904, 2016.

[14] S. Wright, "Evolution in Mendelian Populations.," *Genetics*, vol. 16, no. 2, pp. 97–159, Mar. 1931.

[15] R. Fisher, *The genetical theory of natural selection.* 1930.

[16] M. Kimura, "Evolutionary rate at the molecular level," *Nature*, vol. 217, pp. 624–626, 1968.

[17] M. Lynch, M. S. Ackerman, J.-F. Gout, H. Long, W. Sung, W. K. Thomas, and P. L. Foster, "Genetic drift, selection and the evolution of the mutation rate," *Nat. Rev. Genet.*, vol. 17, no. 11, pp. 704–714, Oct. 2016.

[18] N. H. Barton and B. Charlesworth, "Why sex and recombination?," *Science*

*(80-. ).*, vol. 281, pp. 1986–1990, 1998.

[19] S. P. Otto and T. Lenormand, "Resolving the paradox of sex and recombination," *Nat. Rev. Genet.*, vol. 3, no. 4, pp. 252–261, Apr. 2002.

[20] T. Bedford, S. Riley, I. G. Barr, S. Broor, M. Chadha, N. J. Cox, R. S. Daniels, C. P. Gunasekaran, A. C. Hurt, A. Kelso, A. Klimov, N. S. Lewis, X. Li, J. W. Mccauley, T. Odagiri, V. Potdar, A. Rambaut, D. Wang, X. Xu, P. Lemey, and C. A. Russell, "Global circulation patterns of seasonal influenza viruses vary with antigenic drift," *Nature*, vol. 523, pp. 217–222, 2015.

[21] M. Luksza, M. Lässig, M. Łuksza, and M. Lässig, "A predictive fitness model for influenza.," *Nature*, vol. 507, no. 7490, pp. 57–61, Mar. 2014.

[22] C. A. Suttle, "Viruses in the sea," *Nature*, vol. 437, no. 7057, pp. 356–361, 2005.

[23] D. Paez-Espino, E. A. Eloe-Fadrosh, G. A. Pavlopoulos, A. D. Thomas, M. Huntemann, N. Mikhailova, E. Rubin, N. N. Ivanova, and N. C. Kyrpides, "Uncovering Earth's virome," *Nature*, vol. 536, no. 7617, pp. 425–430, 2016.

[24] A. F. Andersson and J. F. Banfield, "Virus population dynamics and acquired virus resistance in natural microbial communities," *Science (80-. ).*, vol. 320, no. 5879, pp. 1047–1050, 2008.

[25] T. Argov, G. Azulay, A. Pasechnek, O. Stadnyuk, S. Ran-Sapir, I. Borovok, N. Sigal, and A. A. Herskovits, "Temperate bacteriophages as regulators of host behavior," *Curr. Opin. Microbiol.*, vol. 38, pp. 81–87, 2017.

[26] M. A. Beaumont, "Approximate Bayesian Computation in Evolution and Ecology," *Annu. Rev. Ecol. Evol. Syst.*, vol. 41, no. 1, pp. 379–406, Dec. 2010.

[27] P. J. Diggle, R. J. Gratton, and R. J. Grattont, "Monte Carlo Methods of Inference for Implicit Statistical Models," *J. R. Stat. Soc. Ser. B*, vol. 46, no. 2, pp. 193–227, 1984.

[28] D. B. Rubin, "Bayesianly Justifiable and Relevant Frequency Calculations for the Applied Statistician," *Ann. Stat.*, vol. 12, no. 4, pp. 1151–1172, Dec. 1984.

[29] P. Marjoram, J. Molitor, V. Plagnol, and S. Tavare, "Markov chain Monte Carlo without likelihoods.," *Proc. Natl. Acad. Sci. U. S. A.*, vol. 100, no. 26, pp. 15324–8, Dec. 2003.

[30] T. P. Minka, "Expectation Propagation for Approximate Bayesian Inference," 10-Jan-2013. [Online]. Available: https://arxiv.org/abs/1301.2294. [Accessed: 17-Oct-2017].

[31] S. Barthelmé and N. Chopin, "Expectation Propagation for Likelihood-Free Inference," *J. Am. Stat. Assoc.*, vol. 109, no. 505, pp. 315–333, Jan. 2014.

[32] M. a Beaumont, W. Zhang, and D. J. Balding, "Approximate Bayesian computation in population genetics.," *Genetics*, vol. 162, no. 4, pp. 2025–35, Dec. 2002.

[33] M. Sunnåker, A. G. Busetto, E. Numminen, J. Corander, M. Foll, and C. Dessimoz, "Approximate Bayesian computation.," *PLoS Comput. Biol.*, vol. 9, no. 1, p. e1002803, Jan. 2013.

[34] C. P. Robert, J.-M. Cornuet, J.-M. Marin, and N. S. Pillai, "Lack of confidence in approximate Bayesian computation model choice.," *Proc. Natl. Acad. Sci. U. S. A.*, vol. 108, no. 37, pp. 15112–7, Sep. 2011.

[35] H. Strathmann, D. Sejdinovic, S. Livingstone, Z. Szabo, and A. Gretton, "Gradient-free Hamiltonian Monte Carlo with efficient kernel exponential families," in *Advances in Neural Information Processing Systems*, 2015, pp. 955–963.

[36] M. Foll, H. Shim, and J. D. Jensen, "WFABC: A Wright-Fisher ABC-based approach for inferring effective population sizes and selection coefficients from

time-sampled data," *Mol. Ecol. Resour.*, vol. 15, no. 1, pp. 87–98, 2015.

[37] I. Goodfellow, Y. Bengio, and A. Courville, *Deep Learning*. MIT Press, 2016.

[38] S. Sheehan and Y. S. Song, "Deep learning for population genetic inference," *PLoS Comput. Biol.*, vol. 12, no. 3, p. e1004845, 2016.

[39] M. J. Roossinck and E. R. Bazán, "Symbiosis: Viruses as Intimate Partners," *Annu. Rev. Virol.*, vol. 4, no. 1, pp. 123–139, Sep. 2017.

[40] C. Angermueller, T. Pärnamaa, L. Parts, S. Oliver, and O. Stegle, "Deep Learning for Computational Biology," *Mol. Syst. Biol.*, vol. 12, no. 12, p. 878, 2016.

[41] A.-S. Malaspinas, "Methods to characterize selective sweeps using time serial samples: an ancient DNA perspective," *Mol. Ecol.*, vol. 25, no. 1, pp. 24–41, Jan. 2016.

[42] M. Powney, P. Williamson, J. Kirkham, and R. Kolamunnage-Dona, "A review of the handling of missing longitudinal outcome data in clinical trials.," *Trials*, vol. 15, p. 237, Jun. 2014.

[43] E. P. Hong and J. W. Park, "Sample size and statistical power calculation in genetic association studies.," *Genomics Inform.*, vol. 10, no. 2, pp. 117–22, Jun. 2012.

[44] J. D. Watson and F. H. Crick, "Molecular structure of nucleic acids; a structure for deoxyribose nucleic acid.," *Nature*, vol. 171, no. 4356, pp. 737–8, Apr. 1953.

[45] R. Wu, "Nucleotide sequence analysis of DNA.," *Nat. New Biol.*, vol. 236, no. 68, pp. 198–200, Apr. 1972.

[46] F. Sanger, S. Nicklen, and A. R. Coulson, "DNA sequencing with chain-terminating inhibitors.," *Proc. Natl. Acad. Sci. U. S. A.*, vol. 74, no. 12, pp. 5463–7, Dec. 1977.

[47] C. W. Fuller, S. Kumar, M. Porel, M. Chien, A. Bibillo, P. B. Stranges, M. Dorwart, C. Tao, Z. Li, W. Guo, S. Shi, D. Korenblum, A. Trans, A. Aguirre, E. Liu, E. T. Harada, J. Pollard, A. Bhat, C. Cech, A. Yang, C. Arnold, M. Palla, J. Hovis, R. Chen, I. Morozova, S. Kalachikov, J. J. Russo, J. J. Kasianowicz, R. Davis, S. Roever, G. M. Church, and J. Ju, "Real-time single-molecule electronic DNA sequencing by synthesis using polymer-tagged nucleotides on a nanopore array," *Proc. Natl. Acad. Sci.*, vol. 113, no. 19, pp. 5233–5238, 2016.

[48] J. Quick, N. J. Loman, S. Duraffour, J. T. Simpson, E. Severi, L. Cowley, J. A. Bore, R. Koundouno, G. Dudas, A. Mikhail, N. Ouédraogo, B. Afrough, A. Bah, J. H. J. Baum, B. Becker-Ziaja, J. P. Boettcher, M. Cabeza-Cabrerizo, Á. Camino-Sánchez, L. L. Carter, J. Doerrbecker, T. Enkirch, I. G.- Dorival, N. Hetzelt, J. Hinzmann, T. Holm, L. E. Kafetzopoulou, M. Koropogui, A. Kosgey, E. Kuisma, C. H. Logue, A. Mazzarelli, S. Meisel, M. Mertens, J. Michel, D. Ngabo, K. Nitzsche, E. Pallasch, L. V. Patrono, J. Portmann, J. G. Repits, N. Y. Rickett, A. Sachse, K. Singethan, I. Vitoriano, R. L. Yemanaberhan, E. G. Zekeng, T. Racine, A. Bello, A. A. Sall, O. Faye, O. Faye, N. Magassouba, C. V. Williams, V. Amburgey, L. Winona, E. Davis, J. Gerlach, F. Washington, V. Monteil, M. Jourdain, M. Bererd, A. Camara, H. Somlare, A. Camara, M. Gerard, G. Bado, B. Baillet, D. Delaune, K. Y. Nebie, A. Diarra, Y. Savane, R. B. Pallawo, G. J. Gutierrez, N. Milhano, I. Roger, C. J. Williams, F. Yattara, K. Lewandowski, J. Taylor, P. Rachwal, D. J. Turner, G. Pollakis, J. A. Hiscox, D. A. Matthews, M. K. O. Shea, A. M. Johnston, D. Wilson, E. Hutley, E. Smit, A. Di Caro, R. Wölfel, K. Stoecker, E. Fleischmann, M. Gabriel, S. A. Weller, L. Koivogui, B. Diallo, S. Keïta, A.

Rambaut, P. Formenty, S. Günther, and M. W. Carroll, "Real-time, portable genome sequencing for Ebola surveillance," *Nature*, vol. 530, no. 7589, pp. 228–232, Feb. 2016.

[49] L. E. Kafetzopoulou, K. Efthymiadis, K. Lewandowski, A. Crook, D. Carter, J. Osborne, E. Aarons, R. Hewson, J. A. Hiscox, M. W. Carroll, R. Vipond, and S. T. Pullan, "Assessment of metagenomic Nanopore and Illumina sequencing for recovering whole genome sequences of chikungunya and dengue viruses directly from clinical samples," *Eurosurveillance*, vol. 23, no. 50, p. 1800228, Dec. 2018.

[50] M. W. Keller, B. L. Rambo-Martin, M. M. Wilson, C. A. Ridenour, S. S. Shepard, T. J. Stark, E. B. Neuhaus, V. G. Dugan, D. E. Wentworth, and J. R. Barnes, "Direct RNA Sequencing of the Coding Complete Influenza A Virus Genome," *Sci. Rep.*, vol. 8, no. 1, p. 14408, Dec. 2018.

[51] A. L. Greninger, S. N. Naccache, S. Federman, G. Yu, P. Mbala, V. Bres, D. Stryke, J. Bouquet, S. Somasekar, J. M. Linnen, R. Dodd, P. Mulembakani, B. S. Schneider, J.-J. Muyembe-Tamfum, S. L. Stramer, and C. Y. Chiu, "Rapid metagenomic identification of viral pathogens in clinical samples by real-time nanopore sequencing analysis," *Genome Med.*, vol. 7, no. 1, p. 99, Dec. 2015.

[52] N. R. Faria, M. U. G. Kraemer, S. C. Hill, J. G. de Jesus, R. S. Aguiar, F. C. M. Iani, J. Xavier, J. Quick, L. du Plessis, S. Dellicour, J. Thézé, R. D. O. Carvalho, G. Baele, C.-H. Wu, P. P. Silveira, M. B. Arruda, M. A. Pereira, G. C. Pereira, J. Lourenço, U. Obolski, L. Abade, T. I. Vasylyeva, M. Giovanetti, D. Yi, D. J. Weiss, G. R. W. Wint, F. M. Shearer, S. Funk, B. Nikolay, V. Fonseca, T. E. R. Adelino, M. A. A. Oliveira, M. V. F. Silva, L. Sacchetto, P. O. Figueiredo, I. M. Rezende, E. M. Mello, R. F. C. Said, D. A. Santos, M. L. Ferraz, M. G. Brito, L. F. Santana, M. T. Menezes, R. M. Brindeiro, A. Tanuri, F. C. P. dos Santos, M. S. Cunha, J. S. Nogueira, I. M. Rocco, A. C. da Costa, S. C. V. Komninakis, V. Azevedo, A. O. Chieppe, E. S. M. Araujo, M. C. L. Mendonça, C. C. dos Santos, C. D. dos Santos, A. M. Mares-Guia, R. M. R. Nogueira, P. C. Sequeira, R. G. Abreu, M. H. O. Garcia, A. L. Abreu, O. Okumoto, E. G. Kroon, C. F. C. de Albuquerque, K. Lewandowski, S. T. Pullan, M. Carroll, T. de Oliveira, E. C. Sabino, R. P. Souza, M. A. Suchard, P. Lemey, G. S. Trindade, B. P. Drumond, A. M. B. Filippis, N. J. Loman, S. Cauchemez, L. C. J. Alcantara, and O. G. Pybus, "Genomic and epidemiological monitoring of yellow fever virus transmission potential," *Science (80-. ).*, vol. 361, no. 6405, pp. 894–899, Aug. 2018.

[53] P. Mbala-Kingebeni, C.-J. Villabona-Arenas, N. Vidal, J. Likofata, J. Nsio-Mbeta, S. Makiala-Mandanda, D. Mukadi, P. Mukadi, C. Kumakamba, B. Djokolo, A. Ayouba, E. Delaporte, M. Peeters, J.-J. Muyembe Tamfum, and S. Ahuka-Mundeke, "Rapid Confirmation of the Zaire Ebola Virus in the Outbreak of the Equateur Province in the Democratic Republic of Congo: Implications for Public Health Interventions," *Clin. Infect. Dis.*, vol. 68, no. 2, pp. 330–333, Jan. 2019.

[54] S. Hansen, V. Dill, M. A. Shalaby, M. Eschbaumer, S. Böhlken-Fascher, B. Hoffmann, C.-P. Czerny, and A. Abd El Wahed, "Serotyping of foot-and-mouth disease virus using oxford nanopore sequencing," *J. Virol. Methods*, vol. 263, pp. 50–53, Jan. 2019.

[55] S. Theuns, B. Vanmechelen, Q. Bernaert, W. Deboutte, M. Vandenhole, L. Beller, J. Matthijnssens, P. Maes, and H. J. Nauwynck, "Nanopore sequencing as a revolutionary diagnostic tool for porcine viral enteric disease complexes identifies porcine kobuvirus as an important enteric virus," *Sci. Rep.*, vol. 8, no.

1, p. 9830, Dec. 2018.

[56] M. D. Gallagher, I. Matejusova, L. Nguyen, N. M. Ruane, K. Falk, and D. J. Macqueen, "Nanopore sequencing for rapid diagnostics of salmonid RNA viruses," *Sci. Rep.*, vol. 8, no. 1, p. 16307, Dec. 2018.

[57] I. Prazsák, N. Moldován, Z. Balázs, D. Tombácz, K. Megyeri, A. Szűcs, Z. Csabai, and Z. Boldogkői, "Long-read sequencing uncovers a complex transcriptome topology in varicella zoster virus," *BMC Genomics*, vol. 19, no. 1, p. 873, Dec. 2018.

[58] D. Tombácz, I. Prazsák, A. Szűcs, B. Dénes, M. Snyder, and Z. Boldogkői, "Dynamic transcriptome profiling dataset of vaccinia virus obtained from long-read sequencing techniques," *Gigascience*, vol. 7, no. 12, Dec. 2018.

[59] D. Tombácz, Z. Balázs, Z. Csabai, M. Snyder, and Z. Boldogkői, "Long-Read Sequencing Revealed an Extensive Transcript Complexity in Herpesviruses," *Front. Genet.*, vol. 9, p. 259, Jul. 2018.

[60] D. R. Garalde, E. A. Snell, D. Jachimowicz, B. Sipos, J. H. Lloyd, M. Bruce, N. Pantic, T. Admassu, P. James, A. Warland, M. Jordan, J. Ciccone, S. Serra, J. Keenan, S. Martin, L. McNeill, E. J. Wallace, L. Jayasinghe, C. Wright, J. Blasco, S. Young, D. Brocklebank, S. Juul, J. Clarke, A. J. Heron, and D. J. Turner, "Highly parallel direct RNA sequencing on an array of nanopores," *Nat. Methods*, vol. 15, no. 3, pp. 201–206, Jan. 2018.

[61] E. A. Dinsdale, R. A. Edwards, D. Hall, F. Angly, M. Breitbart, J. M. Brulc, M. Furlan, C. Desnues, M. Haynes, L. Li, L. McDaniel, M. A. Moran, K. E. Nelson, C. Nilsson, R. Olson, J. Paul, B. R. Brito, Y. Ruan, B. K. Swan, R. Stevens, D. L. Valentine, R. V. Thurber, L. Wegley, B. A. White, and F. Rohwer, "Functional metagenomic profiling of nine biomes," *Nature*, vol. 452, no. 7187, pp. 629–632, Apr. 2008.

[62] S. Roux, F. Enault, B. L. Hurwitz, and M. B. Sullivan, "VirSorter: mining viral signal from microbial genomic data," *PeerJ*, vol. 3, p. e985, May 2015.

[63] A. A. Pratama and J. D. van Elsas, "The 'Neglected' Soil Virome – Potential Role and Impact," *Trends Microbiol.*, vol. 26, no. 8, pp. 649–662, Aug. 2018.

[64] C. M. Bellas, A. M. Anesio, and G. Barker, "Analysis of virus genomes from glacial environments reveals novel virus groups with unusual host interactions," *Front. Microbiol.*, vol. 6, no. JUL, pp. 1–14, 2015.

[65] C. M. Bellas, A. M. Anesio, J. Telling, M. Stibal, M. Tranter, and S. Davis, "Viral impacts on bacterial communities in Arctic cryoconite," *Environ. Res. Lett.*, vol. 8, no. 4, p. 045021, Dec. 2013.

[66] E. A. G. Schuur and B. Abbott, "Climate change: High risk of permafrost thaw," *Nature*, vol. 480, no. 7375, pp. 32–33, Nov. 2011.

[67] J. Colangelo-Lillis, H. Eicken, S. D. Carpenter, and J. W. Deming, "Evidence for marine origin and microbial-viral habitability of sub-zero hypersaline aqueous inclusions within permafrost near Barrow, Alaska," *FEMS Microbiol. Ecol.*, vol. 92, no. 5, pp. 1–15, 2016.

[68] G. Trubl, N. Solonenko, L. Chittick, S. A. Solonenko, V. I. Rich, and M. B. Sullivan, "Optimization of viral resuspension methods for carbon-rich soils along a permafrost thaw gradient," no. C, pp. 1–24, 2016.

[69] C. M. Bellas and A. M. Anesio, "High diversity and potential origins of T4-type bacteriophages on the surface of Arctic glaciers," *Extremophiles*, vol. 17, no. 5, pp. 861–870, 2013.

## 6 Tables and Legends

| Year | 2013 | 2014 | 2015 | 2016 | 2017 | 2018 |
|---|---|---|---|---|---|---|
| BioRxiv "nanopore" papers | 1 | 13 | 38 | 87 | 170 | 323 |

**Table 1.** The number of papers with the keyword "Nanopore" at the preprint server for biology (BioRxiv) from 2013 to 2018.

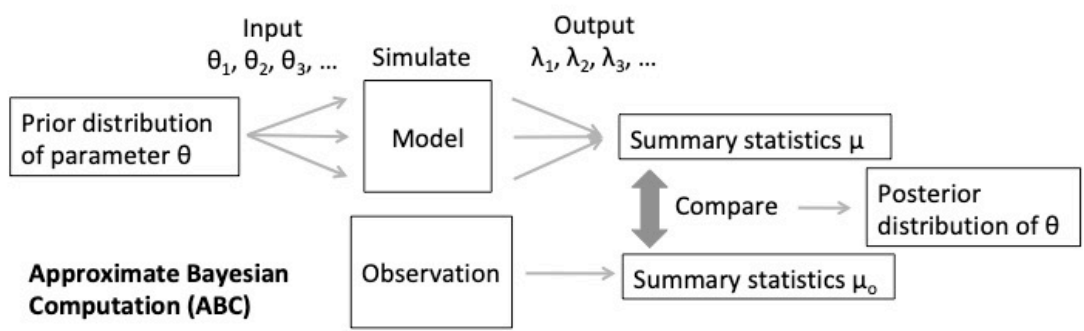

**Figure 1.** Illustration of the Approximation Bayesian Computation (ABC). ABC simulates a model with a set of input parameters and compares its output values to the observation to build a posterior distribution of parameters of interest.

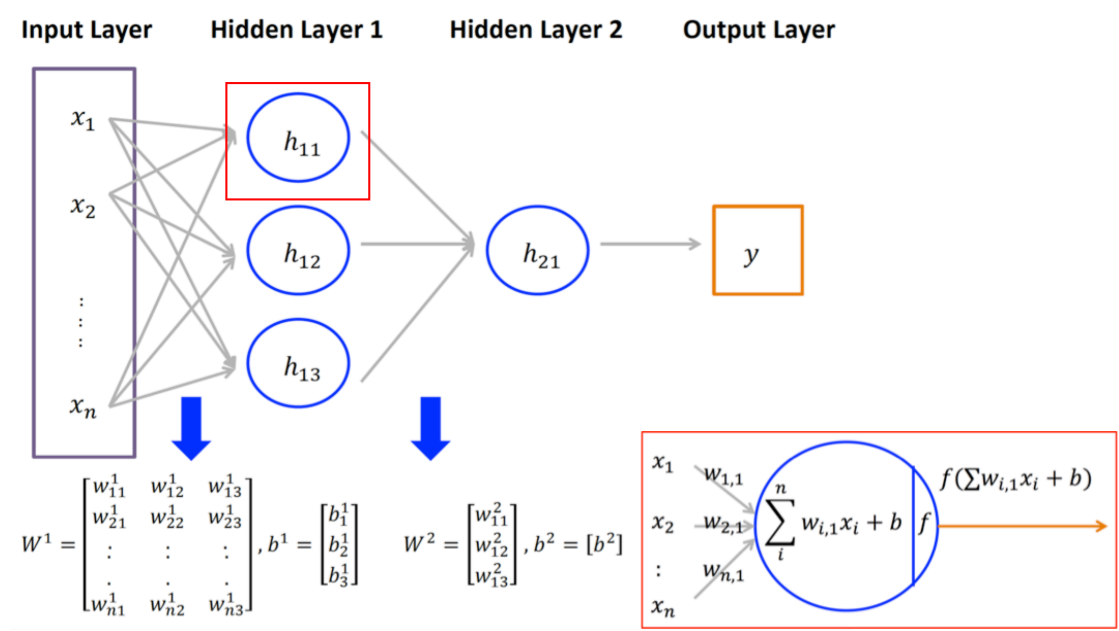

**Figure 2.** Illustration of a fully-connected multilayer perceptron network.

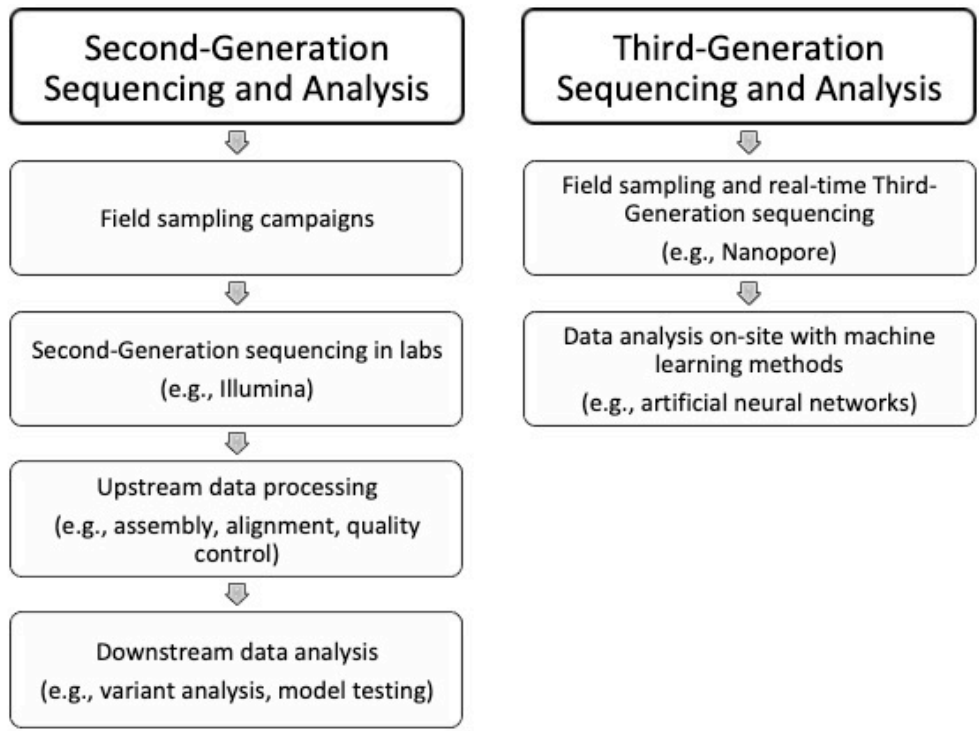

**Figure 3.** Workflow of Second-Generation sequencing and analysis versus workflow of Third-Generation sequencing and analysis.

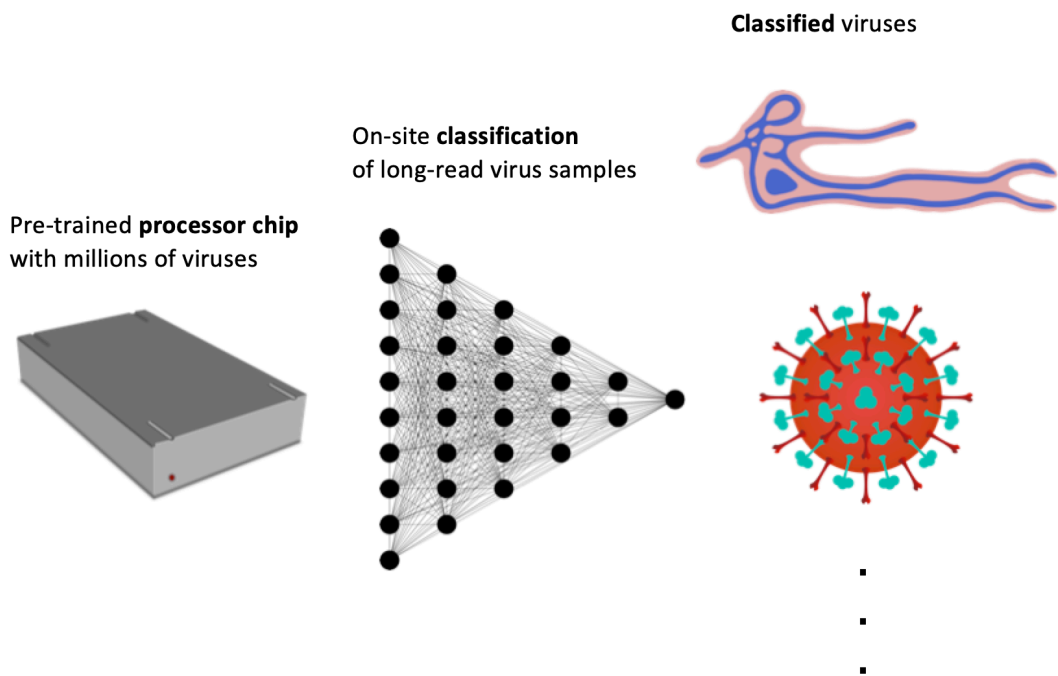

**Figure 4.** Illustration of supervised learning for on-site virus classification with a pre-trained processor chip.

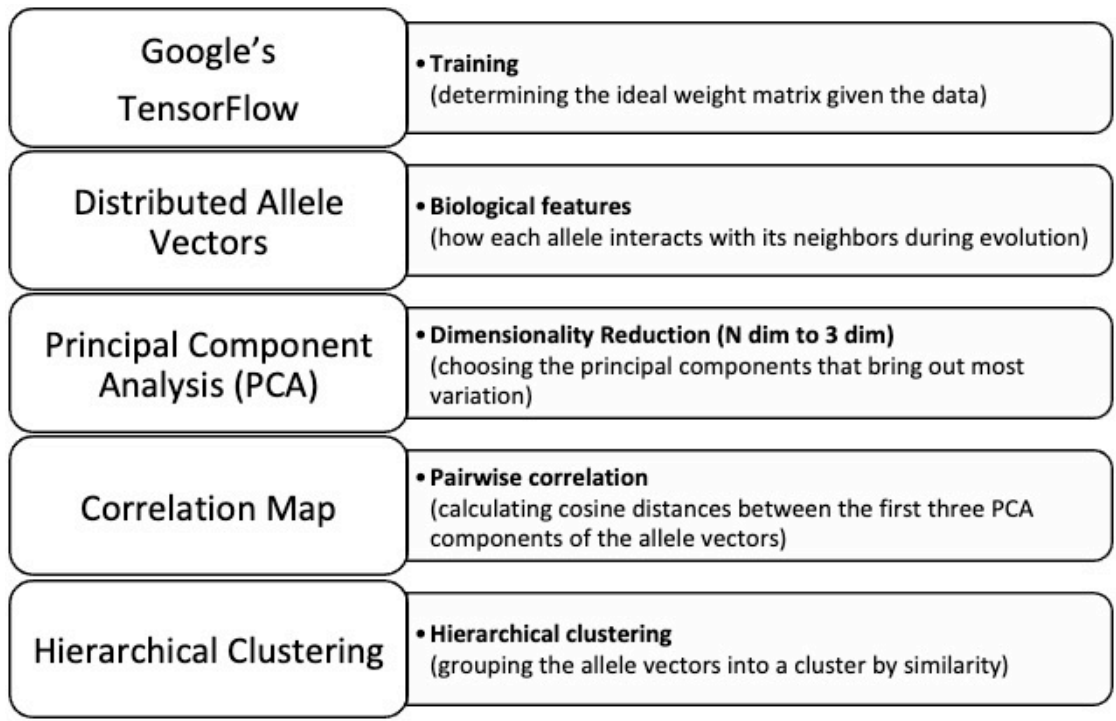

**Figure 5.** Workflow of unsupervised learning of virus genome evolution with the Nucleotide Skip-Gram Neural Network.